\documentclass[12pt]{iopart}

\begin{document}

\jl{6}

\letter{$p$-Branes from Generalized Yang--Mills Theory}

\author{Stefano Ansoldi\dag\footnote[3]{e-mail
		address: \texttt{ansoldi@trieste.infn.it}},
		Carlos Castro\ddag\footnote[4]{e-mail
		address: \texttt{castro@ctsps.cau.edu}},
		Euro Spallucci\dag\footnote[5]{e-mail
		address: \texttt{spallucci@trieste.infn.it}}}

\address{\dag\ Dipartimento di Fisica Teorica
Universit\`a di Trieste, and INFN, Sezione di Trieste}

\address{\ddag\ Center for Theoretical Studies of Physical Systems
Clark Atlanta University, Atlanta, GA.30314}

\begin{abstract}
    We consider the reduced, quenched version of a generalized
    Yang--Mills action in $4k$-dimensional spacetime. This is a
    new kind of matrix theory which is mapped through the
    Weyl--Wigner--Moyal correspondence into a field theory over
    a non-commutative phase space. We show that the ``classical''
    limit of this field theory is encoded into the effective
    action of an open, $(4k-1)$-dimensional, bulk brane enclosed
    by a dynamical, Chern--Simons type, $(4k-2)$-dimensional,
    boundary brane. The bulk action is a pure volume term, while
    the boundary action carries all the dynamical degrees of
    freedom.
\end{abstract}

\pacno{11.17}

\submitted

\maketitle

\noindent
The relation between gauge theories and relativistic extended objects
is one of the most intriguing open
problems currently under investigation
in high energy theoretical physics. Gauge symmetry is the inspiring
principle underlying unification of fundamental forces at the quantum
level, gravity not included. A really unified theory, including
a consistent quantum theory of gravitational phenomena as well, forces
the introduction of relativistic extended objects as the basic building
blocks of matter, space and time. If correct, this picture must be able
to account for the low energy role of gauge symmetry. The presence of
massless vector excitations, carrying Chan--Paton indices
in the massless sector of the open string spectrum,
is a first step towards the answer of this problem in a perturbative
framework.  The recent proposals for a non-perturbative formulation
of string theory in terms of matrices and
$D$-branes \cite{ikkt} provides
further clues in favor of the strings/gauge fields. The problem is
equally difficult to deal from the low energy viewpoint, involving
non-perturbative aspects of gauge theory. Looking for extended
excitations in the spectrum of Abelian gauge
theories is a problem dating
back to the seminal Dirac's work about
strings and monopoles \cite{dirac}.
Recent generalization to higher rank gauge fields has been given in
\cite{noi1}, \cite{noi2}, \cite{luca}.
In the non-Abelian case the problem
is even more difficult because of the interplay with confinement
\cite{antonov}. Thus, it can be dealt
within some appropriate approximation
scheme. Because of the the large value of the gauge coupling constant
standard perturbation theory is not
available and different computational
techniques  have to be adopted. One of the most successful is the
large-$N$ expansion, where $N$ refers
to the number of colors \cite{largen}.
To match Yang--Mills theory and matrix
string theory further approximations
are available, i.e. ``quenching'' and  ``reduction''.
The original $SU(N)$ Yang--Mills field is
replaced by the same field at a single point \cite{reduced},
say $x^\mu=0$ (for a recent review see {\cite{makee}}) and represented
by a unitary $N \times N$ matrix
${\mbox{\boldmath{$A$}}} _{\mu} {}^{i} _{j}$.
Partial derivative operators
are replaced by commutators with a fixed diagonal matrix
${\mbox{\boldmath{$p$}}}_\mu {}^i_j$, playing the role of translation
generator and called the \textit{quenched momentum} {\cite{quench}}.
Accordingly, the covariant derivative becomes
$\rmi {\mbox{\boldmath{$D$}}}_\mu =
\left[ {\mbox{\boldmath{$p$}}}_\mu  + {\mbox{\boldmath{$A$}}}_\mu   ,
\dots \right] $.
Thus, the reduced, quenched, Yang--Mills field  strength is
\[
    {\mbox{\boldmath{$F$}}}_{\mu\nu}{}^i_j\equiv
    \left[\rmi {\mbox{\boldmath{$D$}}}_\mu,
	\rmi {\mbox{\boldmath{$D$}}}_\nu
    \right]^i_j
\]
It has been shown in \cite{bars} that the
dynamics of reduced, quenched,
Yang--Mills theory can be formulated
in the large-$N$ limit in terms of
pure string dynamics.  In a recent paper we have generalized
this result to include bags and membranes in the  spectrum of
$4$-dimensional Yang--Mills theory \cite{hbag}.
In this note, we shall look
for higher dimensional objects fitting
into generalized Yang--Mills theories,
in more than $4$-dimensional spacetime.

Yang--Mills theory admits in $D=4k$, $k=1, 2,\dots$ dimensions
a  generalization  preserving both the canonical dimension of the
gauge field, i.e.
$\left[ {\mbox{\boldmath{$A$}}}_\mu  \right]=( \mathrm{length} )^{-1}$,
and of the coupling constant $\left[  g_{\mathrm{YM}} \right]=1$.
The action we shall use is of the form introduced in \cite{dolant}
\begin{equation}
    \fl S^{\mathrm{GYM}}=-{1\over 2 (2k)! g^2_{\mathrm{YM}}}
    \int d^{4k}x \mathrm{Tr}
	{\mbox{\boldmath{$F$}}}_{[ \mu_1\mu_2}\cdot\dots\cdot
    {\mbox{\boldmath{$F$}}}_{\mu_{2k-1}\mu_{2k}]}
	{\mbox{\boldmath{$F$}}}^{[ \mu_1\mu_2}
    \cdot\dots\cdot
    {\mbox{\boldmath{$F$}}}^{\mu_{2k-1}\mu_{2k}]}
,
\label{gym}
\end{equation}
where ${\mbox{\boldmath{$F$}}}_{\mu_1\mu_2}\equiv \partial_{[\mu_1}
{\mbox{\boldmath{$A$}}}_{\mu_2]}-\left[ {\mbox{\boldmath{$A$}}}_{\mu_1},
{\mbox{\boldmath{$A$}}}_{\mu_2}\right]$,
and the trace operation is over internal indices.
The reason why this particular generalization of the Yang--Mills action
is expected to be of specific interest can be traced back to the original
paper where it was introduced \cite{dolant}. The problem is that the Polyakov
action is conformally invariant for a string but does not preserve such a
fundamental symmetry for higher dimensional objects. The purpose of the authors
in \cite{dolant} was to look for {\it conformally invariant} $\sigma$--models 
in $2n$ dimensional spacetime generalizing the Polyakov action. They introduced
a generalized Yang--Mills gauge theory fulfilling the above requirements and 
reproducing the action (\ref{gym}) when $2n=4k$. In a successive paper the same 
authors identified the instanton solutions of their  generalized Yang--Mills 
model as ``$n$--branes \cite{dt2}. This result was extended also to non-compact
non-linear $\sigma$-models on Anti DeSitter spacetime background in \cite{dt3}. 
From our vantage point, we know that the action (\ref{gym}) with
$k=1$ admits both string and bag solutions in the large--$N$ limit \cite{hbag}.
Thus, we are confident to find $4k-1$--branes in the large--$N$ spectrum
of (\ref{gym}). This type of objects have a ``trivial'' bulk dynamics since
they are spacetime volume filling solutions carrying no transverse degree of 
freedom. On the other hand, a $4k-1$--brane embedded into a $4k$--dimensional
spacetime must be open by definition, and can admit a non--trivial 
$4k-2$--brane as its {\it dynamical boundary}. Thus, we supplement the original 
Dolan Tchrakian action  with a new topological term extending to $4k$ dimension 
the usual Yang--Mills $\theta$-term:

\begin{equation}
    S^{\theta}=-{\theta g^2_{\mathrm{YM}}\over 4\pi^2 4 ^{k}}
    \epsilon^{\mu_1\mu_2 \dots  \mu_{4k-1}\mu_{4k}}
    \int d^{4k}x \mathrm{Tr}   {\mbox{\boldmath{$F$}}}_{[ \mu_1\mu_2}
    \cdot\dots\cdot
    {\mbox{\boldmath{$F$}}}_{\mu_{4k-1}\mu_{4k}]}
 .
\label{atheta}
\end{equation}

The main purpose of this note is to establish a correspondence
between the action $S^{\mathrm{GYM}}+ S^{\theta}$ and some appropriate
brane action. As a first step towards this result we turn the
gauge field actions $S^{\mathrm{GYM}}$ and $S^{\theta}$
into matrix action
through quenching and reduction:
\begin{eqnarray}
    \fl S^{\mathrm{GYM}}
	+
	S^{\theta}\Longrightarrow S^{\mathrm{q, GYM}}_{\mathrm{red.}}
	+
	S^{q, \theta}_{\mathrm{red.}}
\nonumber\\
    \fl \qquad S^{\mathrm{q,  GYM}}_{\mathrm{red.}}
	=
	-{N\over 2 (2k)!  g^2_{\mathrm{YM}}}
    \left( {2\pi\over a}\right)^{4k}
    \mathrm{Tr}   \left[ {\mbox{\boldmath{$D$}}}_{[ \mu_1},
    {\mbox{\boldmath{$D$}}}_{\mu_2} \right]\cdot\dots\cdot
    \left[ {\mbox{\boldmath{$D$}}}_{\mu_{2k-1}},
    {\mbox{\boldmath{$D$}}}_{\mu_{2k}]}\right]
\label{ymqr}
\\
    \times \left[ {\mbox{\boldmath{$D$}}}^{[ \mu_1},
    {\mbox{\boldmath{$D$}}}^{\mu_2} \right]\cdot\dots\cdot
    \left[ {\mbox{\boldmath{$D$}}}^{\mu_{2k-1}},
    {\mbox{\boldmath{$D$}}}^{\mu_{2k}]}\right]
\nonumber \\
    \fl \qquad S^{q,\theta}_{\mathrm{red.}}
	=
	-
	{\theta g^2_{\mathrm{YM}}\over 4\pi^2 4 ^{k}} \left(
    {2\pi\over a}\right)^{4k}
    \epsilon^{\mu_1\mu_2 \dots  \mu_{4k-1}\mu_{4k}}
    \mathrm{Tr}   \left[ {\mbox{\boldmath{$D$}}}_{[ \mu_1},
    {\mbox{\boldmath{$D$}}}_{\mu_2}\right]\cdot\dots\cdot
    \left[ {\mbox{\boldmath{$D$}}}_{\mu_{4k-1}},
    {\mbox{\boldmath{$D$}}}_{\mu_{4k}]}\right]
.
\label{thetaqr}
\end{eqnarray}
One of the most effective way to quantize a
theory where the dynamical variables
are represented by unitary operators is provided
by the  Wigner--Weyl--Moyal
approach \cite{moyal}. A by--product of the
Wigner--Weyl--Moyal quantization
of a Yang--Mills matrix theory is that the 
``classical limit $\hbar \to 0$'' is
just the same as the large-$N$ limit. Once applied to our problem the
Wigner--Weyl--Moyal quantization scheme allows us to write
the unitary matrix ${\mbox{\boldmath{$D$}}}_\mu {}^i_j$
in terms of $2n$ independent matrices
${\mbox{\boldmath{$p$}}}_i$,
${\mbox{\boldmath{$q$}}}_j$, $i, j=1,\dots,n$,
$ 0 \le n \le 4k$ \cite{sochi},
\begin{equation}
    {\mbox{\boldmath{$D$}}}_\mu \equiv {1\over (2\pi)^D}
	\int d^n p\, d^n q
    {\mathcal{A}}_\mu( q, p)
    \exp\left( \rmi q^i  {\mbox{\boldmath{$p$}}}_i +  \rmi p^j
    {\mbox{\boldmath{$q$}}}_j\right)
,
\label{wwm}
\end{equation}
where the operators ${\mbox{\boldmath{$p$}}}_i$, 
${\mbox{\boldmath{$q$}}}^j$
satisfy the Heisenberg algebra
\[
    \left[ {\mbox{\boldmath{$p$}}}_i,
    {\mbox{\boldmath{$q$}}}_j\right]= - \rmi \hbar \delta_{ij}
\]
and $\left( q^i, p^j\right)$ play the role of coordinates in Fourier
dual space. $\hbar$ is the \textit{deformation parameter},
which for historical
reason is often represented by the same symbol as
the Planck constant.\\
The basic idea under this approach is to identify the Fourier
space as the dual of a $2n$-dimensional world manifold of a 
$p=2n-1$ brane.
Consistency requires that the dimension of the
world surface swept by the
brane evolution at most matches the dimension of 
the target spacetime, and
never exceeds it, i.e. $2n \le 4k$.\\
The case $n=1$, $k=1$, describing string sector of large-$N$ QCD,
deserved an in depth investigation \cite{bars}, \cite{pleba}; the case
$n=2$, $k=1$ has been considered in \cite{hbag}.
In this letter, we shall discuss the more general case, $ n = 2k $
and show that it contains a $4k-1$ open brane enclosed by a dynamical boundary.
\\
By inverting \eref{wwm} one gets
\begin{equation}
    {\mathcal{A}}_\mu( q, p)={1\over N}\, \mathrm{Tr}_{{\mathcal{H}}}\,
	\left[
    {\mbox{\boldmath{$D$}}}_\mu
    \exp
	\left( -\rmi {\mbox{\boldmath{$p$}}}_i q^i - \rmi
    {\mbox{\boldmath{$q$}}}_j p^j\,
	\right)
	\right]
,
\label{fasea}
\end{equation}
where $\mathrm{Tr}_{{\mathcal{H}}}$ means the 
sum over diagonal elements with respect
an orthonormal basis in the Hilbert space 
${\mathcal{H}}$ of square integrable
functions on $ R^{4k}$. By Fourier anti--transforming \eref{fasea} one
get the Weyl symbol  ${\mathcal{A}}_\mu( x, y)  $ of the operator
${\mbox{\boldmath{$D$}}}_\mu$:
\[
    {\mathcal{A}}_\mu( x, y)=\int d^n q \, d^n p\, 
    {\mathcal{A}}_\mu( q, p)
    \exp\left( \, i q_i x^i +i p_j y^j\, \right)
.
\]

The above procedure turns the product of two matrices 
${\mbox{\boldmath{$U$}}}$ and
${\mbox{\boldmath{$V$}}}$ into the Moyal, or 
$\ast$-product, of their associated
symbols
\begin{eqnarray}
    {\mbox{\boldmath{$U$}}} {\mbox{\boldmath{$V$}}}
	\longleftrightarrow
    {\mathcal{U}}(\sigma ) \ast {\mathcal{V}}(\sigma)
	\equiv
    \exp\left[ i {\hbar\over 2}\omega^{ab}
    {\partial^2\over \partial \sigma^a \partial \xi^b}
    \right]{\mathcal{U}}(\sigma) {\mathcal{V}}(\xi)
    \vert_{\sigma=\xi}
\nonumber \\
    \sigma^a\equiv \left( x^k, y^l \right)
,
\nonumber
\end{eqnarray}
where $\omega^{ab}$ is the symplectic form
defined over the dual phase space
$(x,y)$.
The introduction of the
non-commutative $\ast$-product allows to express
the commutator between two
matrices ${\mbox{\boldmath{$U$}}}$, ${\mbox{\boldmath{$V$}}}$ as the
\textit{Moyal Bracket} between their
corresponding symbols  ${\mathcal{U}}(x,y)$, ${\mathcal{V}}(x,y)$
\[
    {1\over \hbar}\,  \left[ \, {\mbox{\boldmath{$U$}}},
    {\mbox{\boldmath{$V$}}}\, \right]\longleftrightarrow
    \left\{{\mathcal{U}}, {\mathcal{V}}\right\}_{\mathrm{MB}}
	\equiv
	{1\over \hbar}\,
	\left(\, 
    {\mathcal{U}} \ast {\mathcal{V}} - {\mathcal{V}} \ast {\mathcal{U}}
	\, \right)
    \equiv
	\omega^{ij}\, \partial_i\, {\mathcal{U}}\circ \partial_j \,{\mathcal{V}}
,
\]
where we introduced the $\circ$-product which
corresponds to the ``even'' part of the of the $\ast$-product
{\cite{strach}}.
Once each operator is replaced by its own Weyl symbol,
the trace operation
in Hilbert space turns  into an integration over
a $2D$-dimensional, non-commutative manifold,
because of the ubiquitous presence of the $\ast$ product \cite{madore}:
\[
    {(2\pi)^4\over N^3}
    \mathrm{Tr}_{{\mathcal{H}}} \longmapsto \int d^n x \, d^n y
	\equiv
	\int d^{2n}\sigma
.
\]
The last step of the mapping between matrix
theory into a field model is carried
out through the identification of the
``\textit{deformation parameter}''
$\hbar$ with the inverse of $N$:
\[
    \mbox{``$\hbar$''}\equiv {2\pi\over N}
.
\]
Thus, the large-$N$ limit of the $SU( N )$ matrix theory, where
the ${\mbox{\boldmath{$A$}}}_\mu$ quantum fluctuations freeze,
corresponds
to the   quantum mechanical classical limit, $\hbar\to 0$, of the
WWM corresponding field theory.  From now on, we shall refer to the ``classical
limit'' without distinguishing between the large-$N$ or small-$\hbar$.
In the classical limit the Moyal bracket reproduces the Poisson bracket:
\[
    {\mathcal{F}}_{\mu\nu}\equiv
    \left\{{\mathcal{A}}_\mu, {\mathcal{A}}_\nu\right\}_{\mathrm{MB}}
    \stackrel{\hbar\to0}{\longmapsto}
    {\mathcal{F}}_{\mu\nu}^\infty\equiv  \left\{{\mathcal{A}}_\mu,
    {\mathcal{A}}_\nu\right\}_{\mathrm{PB}}
.
\]

The above formulae are all we need to map the matrix actions
\eref{ymqr} and \eref{thetaqr}  into their Weyl symbols:
\begin{eqnarray}
    \fl W^{\mathrm{GYM}}
    = - {1\over 2 g^2_{\mathrm{YM}} (2k)!}
    \left( {2\pi\over a}\right)^{4k}
    \left( {2\pi\over N}\right)^{2k-4}
\nonumber\\
    \times  \int_\Sigma d^{2n}\sigma
    {\mathcal{F}}_{[ \mu_1\mu_2} \ast \dots
    \ast {\mathcal{F}}_{\mu_{2k-1}\mu_{2k}]}
    \ast {\mathcal{F}}^{[ \mu_1\mu_2} \ast \dots
    \ast {\mathcal{F}}^{\mu_{2k-1}\mu_{2k}]}
\nonumber\\
    \lo{=} {1\over 2 g^2_{\mathrm{YM}} (2k)!}
    \left( {2\pi\over a}\right)^{4k}
    \left( {2\pi\over N}\right)^{2k-4}
\nonumber\\
    \times   \int_\Sigma d^{2n}\sigma
    \left\{
	{\mathcal{A}}_{[\mu_1},{\mathcal{A}}_{\mu_2}
	\right\}_{\mathrm{MB}} \ast \dots
    \ast \left\{{\mathcal{A}}_{\mu_{2k-1}},{\mathcal{A}}_{\mu_{2k}]}
    \right\}_{\mathrm{MB}}
\nonumber\\
    \ast
	\left\{
	{\mathcal{A}}^{[ \mu_1},{\mathcal{A}}^{\mu_2}
	\right\}_{\mathrm{MB}}
    \ast
    \dots  \ast
	\left\{
	{\mathcal{A}}^{\mu_{2k-1}},{\mathcal{A}}^{\mu_{2k}]}
    \right\}_{\mathrm{MB}}
,
\label{wym}
\\
    \fl W^{\theta} = - {\theta g^2_{\mathrm{YM}}\over 4\pi^2 4 ^{k}}
    \left( {2\pi\over a} \right)^{4k}
    \left( {2\pi\over N}\right)^{2k-4}
    \epsilon^{\mu_1\mu_2 \dots  \mu_{4k-1}\mu_{4k}}
    \int_{\partial\Sigma} d^{2n}\sigma {\mathcal{F}}_{[ \mu_1\mu_2}\ast
    \dots  \ast {\mathcal{F}}_{\mu_{4k-1}\mu_{4k}]}
\nonumber\\
    \lo{=} - {\theta g^2_{\mathrm{YM}}\over 4\pi^2 4 ^{k}}
    \left( {2\pi\over a} \right)^{4k}
    \left( {2\pi\over N}\right)^{2k-4}
\nonumber\\
    \times  \epsilon^{\mu_1\mu_2 \dots  \mu_{4k-1}\mu_{4k}}
    \int_{\partial\Sigma} d^{2n}\sigma
    \left\{
	{\mathcal{A}}_{[ \mu_1}, {\mathcal{A}}_{\mu_2}
	\right\}_{\mathrm{MB}}\ast
    \dots \ast
    \left\{
	{\mathcal{A}}_{\mu_{4k-1}}, {\mathcal{A}}_{\mu_{4k}]}
	\right\}_{\mathrm{MB}}
.
\label{wtheta}
\end{eqnarray}

To perform the classical limit of \eref{wym} and \eref{wtheta}
let us rescale the field  ${\mathcal{A}}_\mu$ as
\[
    \left( {2\pi\over N}\right)^{{2k-4\over 4k}}{\mathcal{A}}_\mu
	\longrightarrow
    X_\mu
,
\]
and replace the $\ast$-product
with the ordinary, commutative, product between functions. 
Thus, we obtain

\begin{eqnarray}
    \fl  W_\infty^{\mathrm{GYM}}
    =-{1\over 2g^2_{\mathrm{YM}} (2k)!}
    \left( {2\pi\over a}\right)^{4k}
    \int_{\Sigma} d^{4k}\sigma\,
    \partial_{[\, m_1} X_{\mu_1}\partial_{m_2} X_{\mu_2}
	\cdot\dots\cdot
    \partial_{m_{2k-1}} X_{\mu_{2k-1}}\partial_{m_{2k}} X_{\mu_{2k}\, ]}
\nonumber\\
    \times \partial^{[\, a_1} X^{\mu_1}\partial^{a_2} X^{\mu_2}
	\cdot\dots\cdot
    \partial^{a_{2k-1}} X^{\mu_{2k-1}}\partial^{a_{2k}\, ]}
    X^{\mu_{2k}}
\nonumber\\
    \lo{=} -{1\over 2g^2_{\mathrm{YM}} (2k)!}
    \left( {2\pi\over a}\right)^{4k}
    \int_{\Sigma} d^{4k}\sigma\partial_{[ m_1} X_{\mu_1}\partial_{m_2}
    X_{\mu_2}\cdot\dots\cdot
    \partial_{m_{2k-1}} X_{\mu_{2k-1}}\partial_{m_{2k}]} X_{\mu_{2k}}
\nonumber\\
    \times  \partial^{[ \, m_1} X^{\mu_1}\partial^{m_2} X^{\mu_2}
	\cdot\dots\cdot
    \partial^{m_{2k-1}} X^{\mu_{2k-1}}\partial^{m_{2k}\, ]} X^{\mu_{2k}}
\label{wymln}
\end{eqnarray}
and
\begin{eqnarray}
    \fl W^{\theta}_\infty =
    -{\theta g^2_{\mathrm{YM}}\over 4\pi^2 4 ^{k}}
    \left( {2\pi\over a} \right)^{4k}
    \epsilon^{\mu_1\mu_2 \dots  \mu_{4k-1}\mu_{4k}}
    \int_{\partial\Sigma} d^{4k-1}s\, \epsilon^{m_2\dots m_{4k}}
    X_{\mu_1} \partial_{m_2} X_{\mu_2}\cdot\dots\cdot
    \partial_{m_{4k-1}}X_{\mu_{4k}}
\nonumber\\
    \lo{=}  -{\theta g^2_{\mathrm{YM}}\over 4\pi^2 4 ^{k}}
    \left( {2\pi\over a} \right)^{4k}
    \epsilon^{\mu_1\mu_2 \dots  \mu_{4k-1}\mu_{4k}}
    \int_{\partial\Sigma} d^{4k-1}s\,
    X_{\mu_1} \left\{X_{\mu_2} ,\dots
    , X_{\mu_{4k-1}}\right\}_{\mathrm{NPB}}
.
\label{wthetaln}
\end{eqnarray}

The two actions \eref{wymln} and
\eref{wthetaln} are the main results of
this note and the discussion of their physical
meaning will end this letter.\\
$W^{\theta}_\infty$ is the action for the \textit{Chern--Simons} $(4k-2)$-brane,
investigated in \cite{zaikov}. This kind of  $p$-brane is a dynamical object
with very interesting properties following from its topological origin
\cite{neeman}, e.g. the presence of the generalized Nambu--Poisson bracket
$\left\{X_{\mu_2} ,\dots, X_{\mu_{4k-1}}\right\}_{\mathrm{NPB}}$ which
suggests a new formulation of both classical 
and quantum mechanics for this
kind of object \cite{dito}.\\
To identify the kind of physical object described by
$W_\infty^{\mathrm{GYM}}$ we need to recall the
conformally invariant, $4k$-dimensional,  $\sigma$-model  action
introduced in \cite{dt}, \cite{dt2}:
\begin{eqnarray}
    \fl S_{4k-1}=-{1\over (2k)!}T_{4k-1}\int  d^{4k}\sigma
    \sqrt h\, h^{m_1 n_1}\cdot\dots\cdot  h^{m_{2k} n_{2k}}
\nonumber\\
    \times  \
	\partial_{[\, m_1}  X^{\mu_1}
	\cdot\dots\cdot
	\partial_{m_{2k}\, ]}
    X^{\mu_{2k}}
    \partial_{[\, n_1}  X^{\nu_1}
	\cdot\dots\cdot
	\partial_{n_{2k}\, ]}  X^{\nu_{2k}}
    \eta_{\mu_1 \nu_1}
	\cdot\dots\cdot
	\eta_{\mu_{2k} \nu_{2k}}
,
\label{brane}
\end{eqnarray}
where $X^\mu(\sigma)$ are the $p$-brane coordinates in target
spacetime;
$T_{4k-1}$ is a constant with dimensions of energy per unit
$(4k-1)$-volume, or $\left[T_{4k-1}\right]= ( \mathrm{length} )^{-4k}$;
$h_{mn}( \sigma)$ is an auxiliary metric tensor providing
reparametrization invariant over the $p$-brane world volume. For the
sake of simplicity, we assumed the target spacetime to be flat and
contracted the corresponding indices, i.e. $\mu_1, \mu_2, \dots ,
\nu_1, \nu_2, \dots $, by means of a Minkowski tensor. By solving
the classical field equation $\delta  S_{4k-1}/\delta h_{mn}=0$ one can
write  $h_{mn}$ in terms of the induced metric $G_{mn}=\partial_m X^\mu
\partial_n X_\mu$ and recover the Dirac--Nambu--Goto form of the
action of \eref{brane} and identify $T_{4k-1}$ with the
brane tension.  If we break the reparametrization invariance of
the action \eref{brane}, by choosing a conformally flat world
metric
\[
    h_{mn}=\exp\left\{2\Omega(\sigma)\right\} \eta_{mn}
,
\]
the resulting, volume preserving diffeomorphism invariant action, 
is just \eref{wymln} with a tension given by
\begin{equation}
    T_{4k-1}= {1\over g^2_{\mathrm{YM}}}
	\left( {2\pi\over a} \right)^{4k}
.
\label{tens}
\end{equation}
Thus, the large-$N$ limit of the generalized Yang--Mills theory
\eref{gym} describes a \textit{bag--like}, vacuum domain, or 
$(4k-1)$-brane,
characterized by a tension \eref{tens}. Being embedded into a
$4k$-dimensional target spacetime the bag has no transverse, dynamical,
degrees of freedom, i.e. it is a pure volume term. The whole dynamics
is confined to the boundary in a way which seems to saturate the
holographic principle \cite{holo}.\\
We can summarize the results we have obtained in the ``flow chart''
below.

\begin{center}
    \begin{tabular}{|ccccccc|}
 	& & & & & & \\
    $A_\mu{}^i_j\left( x \right)$
 	&
 	$\Rightarrow$
 	&
 	${\mbox{\boldmath{$A$}}}_\mu{}^i_j$
    &
	$\mapsto$
	&
	${\mathcal{A}}_\mu\left( \sigma \right)$
    &
	$\hookrightarrow$
	&
	$X_\mu\left( \sigma \right)$
	\\ & & & & & & \\
    gauge field
	&
	$\Rightarrow$
	&
	matrix
	&
	$\mapsto$
    &
	Weyl symbol
	&
	$\hookrightarrow$
	&
	brane coordinate
	\\ & & & & & & \\
    $S^{\mathrm{GYM}}$
	&
	$\Rightarrow$
	&
	$S^{\mathrm{q, GYM}}_{\mathrm{red.}}$
	&
	$\mapsto$
	&
	$W^{\mathrm{GYM}}$
    &
    $\hookrightarrow$
	&
	$S^{p=4k-1}_{\mathrm{DT}}=$volume term
	\\ & & & & & & \\
	$S^\theta$
	&
	$\Rightarrow$
	&
	$S^{q, \theta}_{\mathrm{red.}}$
	&
	$\mapsto$
	&
	$W^\theta$
    &
    $\hookrightarrow$
	&
	$S^{p=4k-2}_{\mathrm{CS}}=$boundary term
	\\ & & & & & & \\
\end{tabular}
\end{center}
The various arrows respectively represent:
\begin{description}
	\item[$\Rightarrow$] quenching and zero volume limit;
	\item[$\mapsto$] Weyl--Wigner--Moyal mapping;
	\item[$\hookrightarrow$] large-$N$ limit.
\end{description}
Through all these operations we transformed the generalized Yang--Mills theory,
described by \eref{gym} and \eref{atheta} into an \textit{effective theory} 
for higher dimensional, $(4k-1)$-dimensional, vacuum domain of large-$N$,
generalized Yang--Mills theory, bounded by a $(4k-2)$-dimensional
Chern--Simons brane. We conclude this letter with some remarks about a 
$D$--brane interpretation of our result. It has been proved that an
$SU(\,N\,)$ Yang--Mills theory, dimensionally reduced from $d=10$ to 
$d=p+1$, encodes the low energy dynamics of a stack of $N$ coincident
Dirichlet $p$--branes \cite{dbrane}. This physical interpretation follows from
 the  matrix, non--commuting, features shared by Yang--Mills
fields and $D$--branes coordinates. This basic feature is encoded into the 
generalized action (\ref{gym}) as well. We have only to consider that the
 the specific form of the action (\ref{gym}) is spacetime
dimension dependent. However, if we give up conformal invariance in arbitrary
number of spacetime dimensions and stick to the $4k$--type form of the
Dolan Tchrakian  Lagrangian, we can look at the action (\ref{gym}) as the
result of a dimensional reduction from $d>4k$ to $4k$. Then, $D$--brane 
picture can be applied again. Matching this picture with the end results of our
procedure, we conclude that the low energy effective action of a large number of
coincident $4k-1$  Dirichlet branes can be approximated the action
(\ref{wymln}). This conjecture deserves further investigation.

\end{document}